\documentclass[fleqn,useAMS,usenatbib]{mn2e}
\usepackage{graphicx}
\usepackage{amssymb}      
\usepackage{color}
\usepackage{pstricks}
\usepackage{amsmath}
\usepackage{latexsym}
\newcommand{\be}{\begin{equation}}
\newcommand{\ee}{\end{equation}}
\newcommand{\dd}{{\rm d}}
\newcommand{\lp}{\left(}
\newcommand{\rp}{\right)}

\title[Mock catalogues for the REFLEX II sample]{\textbf{Constructing mock catalogues for the REFLEX II galaxy cluster sample}}
\author[Balaguera-Antol\'{\i}nez, S\'anchez, B\"ohringer \& Collins]{\parbox{\textwidth}{
A. Balaguera-Antol\'{\i}nez$^{1,2}$\thanks{E-mail: abalan@astro-uni.bonn.de}, 
Ariel G. S\'anchez$^{2}$, 
H. B\"ohringer$^{2}$,
C. Collins$^{3}$
}\\\\
$^{1}$Argelander Institute f\"ur Astronomie, Auf dem H\"ugel 71, D-53121 Bonn, Germany\\
$^{2}$Max Planck Institute f\"ur Extraterrestrische Physik, D-85748, Garching, Germany\\
$^{3}$Astrophysics Research Institute, Liverpool John Moores University, Birkenhead, Wirral CH41 1LD, U.K.\\
}

\begin{document}
\pagerange{\pageref{firstpage}--\pageref{lastpage}} \pubyear{2012}
\maketitle

\label{firstpage}
                  
\begin{abstract}
We describe the construction of a suite of galaxy cluster mock catalogues from N-body simulations, based on the properties of the new ROSAT-ESO Flux-Limited X-Ray (REFLEX II) galaxy cluster catalogue. Our procedure is based on the measurements of the cluster abundance, and involves the calibration of the underlying scaling relation linking the mass of dark matter haloes to the cluster X-ray luminosity determined in the \emph{ROSAT} energy band $0.1-2.4$ keV. In order to reproduce the observed abundance in the luminosity range probed by the REFLEX II X-ray luminosity function ($0.01<L_{X}/(10^{44}{\rm erg}\,{\rm s}^{-1}h^{-2})<10$), a mass-X ray luminosity relation deviating from a simple power law is required. We discuss the dependence of the calibration of this scaling relation on the X-ray luminosity and the definition of halo masses and analyse the one- and two-point statistical properties of the mock catalogues. Our set of mock catalogues provides samples with self-calibrated scaling relations of galaxy clusters together with inherent properties of flux-limited surveys. This makes them a useful tool to explore different systematic effects and statistical methods involved in constraining both astrophysical and cosmological information from present and future galaxy cluster surveys.
\end{abstract}


%===================================================================================================================================================
\section{Introduction}
Mock catalogues have become a key tool to assess the statistical methods employed in the analysis of redshift galaxy surveys, mainly concerning the study of the large-scale structure of the Universe \citep[e.g][]{cole2df,cabre, ariel_last,percival_last,norberg}. Such tools have been usually constructed following two methods, the first being based on log-normal realizations of the matter density field \citep{coles_jones}, and the second based on the use of N-body simulations. The log-normal catalogues can be produced in a shorter time period, and can be designed such that the final suite of mock observations share the same clustering properties as that of real data \citep[e.g][]{pvp,cole2df}. On the other hand, N-body simulations have the advantage of properly characterising the dynamical evolution and the growth of structures in the non-linear regime, providing a tool to test theoretical predictions for the abundance of dark matter haloes \citep[e.g][]{warren,tinker,crocce_mice_abu} as well as on the non-linear evolution of gravitational clustering of matter, dark matter haloes and galaxies \citep[e.g][]{evrard,millenium,angulo,lasdamas,cai_pans,horizon}. In the context of galaxy clusters surveys, the combination of cosmological N-body simulations, a precise understanding of the survey selection function and a set of cluster-mass proxies is required for the construction of mock catalogues containing the relevant structural an dynamical properties of these objects, such as their masses, X-ray luminosities, spectral temperatures, fluxes and peculiar velocities,
among others. This provides a control sample to analyse the possible systematic effects in the determination of scaling relations and/or cosmological parameters, together with the possibility to assess the capability of future galaxy cluster surveys (e.g {\it eROSITA}\footnote{http://www.mpe.mpg.de/erosita/}) to push the constraints on the latter to higher precision
(e.g \citet[][]{vik, pillepich}).

In the context of cosmology probed by the clustering of galaxy clusters, previous analyses
\citep[e.g][]{ret,reflex1} used cosmological N-body simulations in order
to determine statistical properties such as covariance matrices associated to different measurements of abundance and clustering. On smaller scales, hydro-N-body simulations
are widely used to explore the influence of non-thermal processes such as AGN and Supernova feedback on the cluster scaling relations \citep[e.g][]{evrard,borgani05,stanek_last,short}. 
The implementation of the baryonic physics required to model these
astrophysical processes in large cosmological simulations is
challenging and it would demand extremely large computational resources. This leads to the necessity to implement semi-analytic models, which in the case of galaxy clusters refers to the
cluster scaling relations, such that the mock catalogues match the observed properties of these objects.

This is the second of a series of papers dedicated to the analysis of the new REFLEX II galaxy cluster catalogue. A first paper (\cite{balaguera}, hereafter Paper I) was focused on the clustering analysis of this new sample through the power spectrum. In that work, a set of mock catalogues was used to determine the covariance matrix of the cluster power spectrum and thereby assess the statistical methods implemented in that analysis. In this paper we describe the construction of that suite of mock catalogues, which have been designed to reproduce basic properties of the REFLEX II sample (e.g survey geometry, selection function, cluster abundance, etc.).
A key ingredient in the construction of realistic cluster mock catalogues using N-body simulations is the assignment of luminosities to the dark matter haloes.

Owing to the fact that the one-point statistics (e.g abundance) of galaxy clusters (as well as
their clustering) are sensitive to the underlying mass $M$-X ray luminosity $L_{X}$ relation (hereafter $M$-$L_{X}$) \citep[e.g][]{voit}, it is important to have a
scaling relation consistent with both the observed cluster abundance and a halo mass-function in N-body simulations.
Accordingly, instead of assigning luminosities using scaling relations found in the literature (see for instance the scaling relations obtained by \citet[][]{lm0,stanek,mau,pratt,mantzI}), we calibrate the $M$-$L_{X}$
based on the X-ray luminosity function (XLF hereafter) of the REFLEX II sample. We discuss the main features of the scaling relation, though it is not the scope of this
paper to develop a deep analysis on this topic.

We also briefly describe the two-point statistics of our mock catalogues (broadly described in Fourier space in
Paper I) and show how these can be used to determine covariance matrices in order to develop constraints on cosmological parameters based on likelihood analysis
(S\'anchez et al., in preparation). The worth of our set of mock catalogues is reflected in their physical and observational content. These are translated to the covariance
matrix of the cluster power spectrum (or correlation function) and its capability to properly account for non-linear evolution of the matter density field, dependencies of clustering strength with the underlying scaling relations and systematics effects introduced by the survey selection function. In that sense, the set of REFLEX II mock
catalogues introduced in this paper represents an improvement in the analysis of this sample, compared to previous works \citep[e.g][]{collins_cf,reflex1} based on the
REFLEX sample \citep{hb_catalogo}.  

The outline of this paper is as follows. In \S~\ref{sec:data} we briefly introduce the REFLEX II galaxy cluster catalogue and in \S~\ref{subsec:xlf} we describe the measurement and parameterisation of the REFLEX II XLF. In \S~\ref{sec_mocks} we introduce the N-body simulations and discuss the procedure we follow to calibrate
the underlying $M$-$L_{X}$ relation (\S~\ref{sec:mla} and \S~\ref{sec:results}). In \S~\ref{sec_properties} we
discuss some of the most relevant properties of the mock catalogues. Finally, we end with our main conclusions in \S~\ref{sec_conclusions}.

We adopt a fiducial cosmological model based on a flat $\Lambda$CDM Universe with a matter energy density parameter
of $\Omega_{\rm mat}=0.25$, a dimensionless Hubble parameter $h=0.7$ (in units of $100 \,{\rm km}\,{\rm s}^{-1} {\rm Mpc}^{-1}$) and a rms of mass fluctuations of $\sigma_{8}=0.773$. Unless otherwise stated, redshifts, distances, fluxes, luminosities,
power spectra and mass functions are calculated with this set of cosmological parameters. Throughout this paper we always refer to the rest-frame X-ray luminosity in the \emph{ROSAT} energy band $0.1-2.4$ keV determined under the
assumption of spherical overdensities (SO hereafter) characterised by $\Delta=500$ (i.e, the mean matter
density of the cluster is $500$ times the critical density of the Universe). We will frequently refer to 
\emph{a set of realizations}, namely, the N-body simulations of dark matter haloes. We warn the reader not to confuse these with the \emph{mock catalogues}, which refers to the set of REFLEX II-like catalogues.

\begin{figure}
\includegraphics[width=8.3cm]{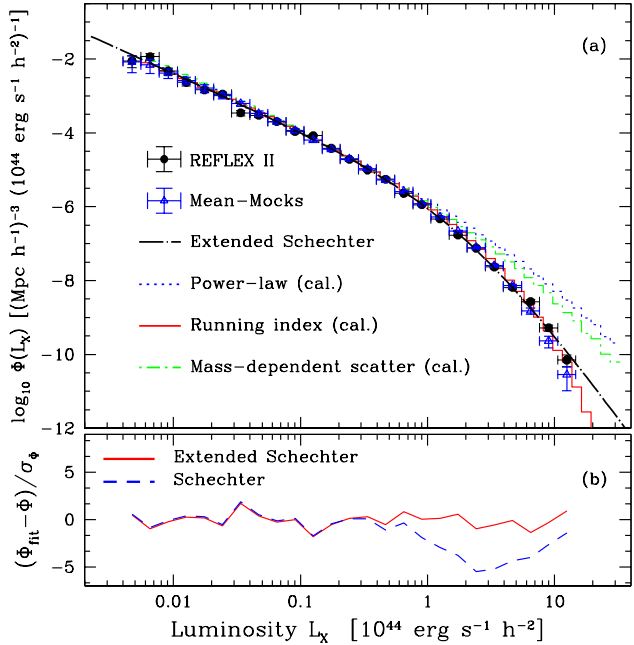}
\caption{(a) The filled points with error bars correspond to the measurement of the REFLEX II X-ray luminosity function $\Phi(L_X)$.
The black dot-dashed line represents the best fitting extended-Schechter function, defined in Eq.~(\ref{XLFe}), to
this measurement. The red solid, blue-dotted and green dot-dashed histograms represent the mean luminosity distribution determined
from the ensemble of $50$ realizations of the L-BASICC II simulation using haloes with X-ray luminosities assigned 
according to i) a $M$-$L_{X}$ relation given by Eq.~(\ref{m_lum_relation}) (running index) with the set  $(a,b,c)$ calibrated according to \S~\ref{sec:mla}, ii) a power-law $M$-$L_{X}$ relation and iii) a power-law with mass-dependent scatter (\S~\ref{sec:phys}), respectively. The triangles represent the mean luminosity function from the final 
set of $100$ mock catalogues. Panel (b) shows the difference between the measurements and the Schechter and extended Schechter fits in units of the standard deviation.}\label{lcasos} 
\end{figure}
\section{The REFLEX II catalogue}\label{sec:data}

\subsection{The REFLEX II sample}
\label{sec:sample}

The REFLEX catalogue is based on the ROSAT All-Sky Survey \citep[RASS,][]{truemper}, where galaxy cluster candidates are detected as X-ray sources
coinciding with galaxy concentrations in optical sky survey images (B\"ohringer et al., in preparation). The final identifications and redshift measurements come
from complementary follow-up observations as described by \citet{guzzo_optical}, yielding spectroscopic redshifts for $860$ clusters 
with flux limit of $1.8 \times 10^{-12}\, {\rm erg}\,{\rm s}^{-1}{\rm cm}^{-2}$ in the \emph{ROSAT} energy band and approximately $6$ per-cent incompleteness in
redshift follow-up. The missing redshifts are currently being obtained though observations at La Silla.  
While a detailed description of the construction of the catalogue and the selection function will be given in forthcoming papers (B\"ohringer et al. in prep.), Paper I
presents a brief summary of the main steps followed in the derivation of the cluster parameters and the characterisation of the survey selection function. 
Here we briefly summarize the aspects that are more relevant to the construction of the mock catalogues.

In our analysis we used the X-ray luminosities corrected for missing flux in the \emph{ROSAT} energy band. The error in the measurement of the flux varies
through the REFLEX II sample, depending on parameters such as the observed flux and the exposure time associated with each cluster. %In terms of the X-ray luminosity, the distribution of the luminosity errors can be well approximated by a gaussian distribution, although the width might depend upon the quantities mentioned above.  
In order to keep the analysis simple, for the construction of the mock catalogues we adopted a normal distribution with a width $\sigma_{L_{X}}=0.2L_{X}$ ($20$ per cent error in
the X-ray luminosity), which fairly characterizes the flux errors in the REFLEX II sample.

The REFLEX II sensitivity map can be described by dividing the surveyed area ($13924$ deg$^{2}$ in the southern hemisphere) into $N_{\rm pix}=13952$ pixels of approximately 
equal area of $1\,{\rm deg}^{2}$. Depending on its exposure time, the limiting luminosity $L_X^{\rm lim}(z)$ for each pixel was tabulated in the range $0\leq z \leq 0.8$ assuming
a nominal flux limit. This defines the REFLEX II selection function.

\subsection{The REFLEX II X-ray luminosity function}
\label{subsec:xlf}
As will be described in section~\ref{sec:mla}, a key ingredient for the construction of the mock catalogues is the X-ray luminosity function
of the REFLEX II sample. We measured the XLF by means of the standard $1/V_{\rm max}$
estimator \citep{felten}. The resulting X-ray luminosity function, shown by the filled points with error bars in panel (a) of Fig.~\ref{lcasos}, can be well described by   
what we call an {\it extended-Schechter function:}
\be
\Phi(L){\rm d} L=n_{0}\left(\frac{L}{L_{\star}}\right)^{-\alpha}{\rm e}_{q}\lp-\frac{L}{L_{\star}}\rp\,{\rm d}\left(\frac{L}{L_{\star}}\right).
\label{XLFe}
\ee
Here $n_{0}$ determines the overall normalisation, $\alpha$ characterises the slope of the XLF in the low-luminosity regime and $L_{\star}$ marks the transition from power-law behaviour at low luminosities to the exponential fall-off at high luminosities. The function ${\rm e}_{q}(x)$ is the so-called $q$-exponential distribution
\citep{tsallis}, defined as 
\be
{\rm e}_{q}(x)=
\begin{cases}
e^{x}& q=1, \\
(1+x(1-q))^{1/(1-q)}&  q\neq 1.\\
\end{cases}
\ee
The best-fitting parameters $(n_{0},\alpha,L_{\star},q)$ to the REFLEX II X-ray luminosity function were determined by implementing a Markov-Chain Monte Carlo
(hereafter MCMC) algorithm \citep[e.g][]{neal_mcmc}. These are $\alpha=1.54\pm 0.06$, $L_{\star}=(0.63\pm 0.15)\times 10^{44}{\rm erg}\, {\rm s} \, h^{-2}$, $n_{0}=(4.08\pm 0.82)\times 10^{-6}({\rm Mpc}\, h^{-1})^{-3}$, and $q=1.31\pm 0.03$, where the quoted errors denote the $68$ per cent confidence levels in each parameter.
The best fitting extended-Schechter function corresponding to these parameter values is shown by the dot-dashed line in panel (a) of Fig.~\ref{lcasos}. Panel (b) of the same figure shows the difference between the measured XLF and the best-fit averaged in the corresponding luminosity bins, in units of the standard deviation. For comparison, in the same panel we show the performance of the Schechter function\footnote{The best-fitting parameters of the Schechter function ($q=1$) are $\alpha=1.71\pm 0.07$, $L_{\star}=(1.51\pm 0.15)\times 10^{44}{\rm erg}\, {\rm s} \, h^{-2}$ and $n_{0}=(1.29\pm 0.91)\times 10^{-6}({\rm Mpc}\, h^{-1})^{-3}$}, showing that such parameterisation cannot fully characterize the REFLEX II XLF, especially at its bright tail.

In order to carry out the statistical analysis using the information of the XLF, we assume a diagonal covariance matrix $C^{\Phi}_{ij}=\sigma^{2}_{\Phi}(L_{i})\delta^{K}_{ij}$, with a variance given by the Poisson standard deviation $\sigma_{\Phi}(L_{i})=\Phi(L_{i})/\sqrt{N_{i}}$, where $N_{i}$ is the number of clusters in the $i-$th luminosity bin. This assumption will be also used within the analysis presented in \S~\ref{sec:nb}, and represents a good description of the covariance matrix of the REFLEX II XLF, as will be shown in \S~\ref{sec_properties}.

\section{Construction of the REFLEX II mock catalogues}
\label{sec_mocks}

\subsection{The N-body simulations}
\label{sec:nb}

Our mock catalogues are based on the Low resolution Baryon Acoustic Simulations at ICC (hereafter L-BASICC II) N-body simulations.
A detailed description of these simulations can be found in \citet{angulo} and \citet{ariel_bao}.
These correspond to 50 realizations of the same flat $\Lambda$CDM model characterised by a matter energy-density
parameter $\Omega_{\rm mat}=0.237$, a baryon energy-density $\Omega_{\rm ba}=0.046$, a dimensionless Hubble parameter $h=0.73$, a
rms of mass fluctuations $\sigma_{8}=0.773$ and a scalar spectral index $n_{s}=0.997$. Each of the L-BASICC II simulations follows the dark matter distribution using $448^3$ particles
over a comoving box of side $1.34\,{\rm Gpc}\,h^{-1}$, comparable to the Hubble Volume Simulation \citep{evrard}. We used the halo samples at $z=0$ constructed by means of a friends-of-friends (hereafter FOF) algorithm with a linking lenght of $b=0.2$. 

Panel (a) of Fig.~\ref{abundance_2} shows the mean halo mass function $n(M)$ from the ensemble of simulations. 
Due to the resolution of these simulations it is only possible to identify halos with masses $M>1.7\times 10^{13}M_{\odot}h^{-1}$ (corresponding to ten dark
matter particles), thus probing only the high-mass tail of the halo mass function, which is very well described by the fitting formula of \citet{jenkins}
(shown by the solid line in panel (a) of Fig.~\ref{abundance_2}). Note that the cosmological model of the L-BASICC II simulations is slightly different
from our fiducial cosmology. We check that the results shown in this paper are not qualitatively modified due to these differences in cosmological parameters
(see \S~$3.5$).

Even though the accuracy of the calibration of the parameters in the different recipes for the mass function of dark matter haloes has been pushed to the
$5$ to $10$ per cent level \citep[e.g][]{jenkins,warren,tinker}, it has been shown that the inclusion of baryons in the N-body simulations increases this
uncertainty to $\sim 15$ per cent \citep{stanek_bar}, especially in the mass range probed by the L-BASICC II simulations. Such uncertainties must in principle
be taken into account when comparing theoretical predictions with observations on cluster counts/abundances. Although we have not accounted for these uncertainties
in the results presented in this paper, we checked that our conclusions do not vary significantly when these are included by allowing up to a $20$ per cent uncertainty
in the normalisation of halo mass function.

Similarly, it has been shown \citep[e.g][]{tinker,lukic,more} that the FOF halo finder systematically overestimates the masses of dark matter haloes, compared to the masses of the same objects found within simulations at higher resolution. This can be a potential systematic effect that might affect the shape of the scaling relations linking the FOF halo masses to cluster observables, in our case, the X-ray luminosity. Using the recent analysis of \citet{more}, we have verified that such corrections in the halo masses do not modify our results and conclusions (see  \S~\ref{sec:phys}).

\subsection{X-ray luminosity assignment}
\label{sec:mla}

The first step in the construction of the REFLEX II mock catalogues from the N-body simulations is the assignment of X-ray luminosities to the dark
matter halos. We do this following a $M$-$L_{X}$ relation which we calibrate in order to reproduce the observed X-ray luminosity function of the REFLEX II sample. 
The information of the scaling relation $M$-$L_{X}$ is encoded in the function $p(L_{X}|M,z)$, specifying the probability of a cluster to have assigned a X-ray luminosity $L_{X}$ conditional to its mass $M$ at redshift $z$. Given the abundance of dark matter haloes $n(M,z)$, the X-ray luminosity function at a given
redshift $z$ is given by:
\be\label{eq:flum_theo}
\Phi(L_{X},z)=\int_{0}^{\infty}\, n(M,z)\,p(L_{X}|M,z)\,{\rm d} M.
\ee
In this way, given the halo abundance from the N-body simulations, Eq.~(\ref{eq:flum_theo}) allows us to calibrate the
$M$-$L_{X}$ relation in order to reproduce the measured XLF.

\begin{figure}
\includegraphics[width=8.5cm, angle=0]{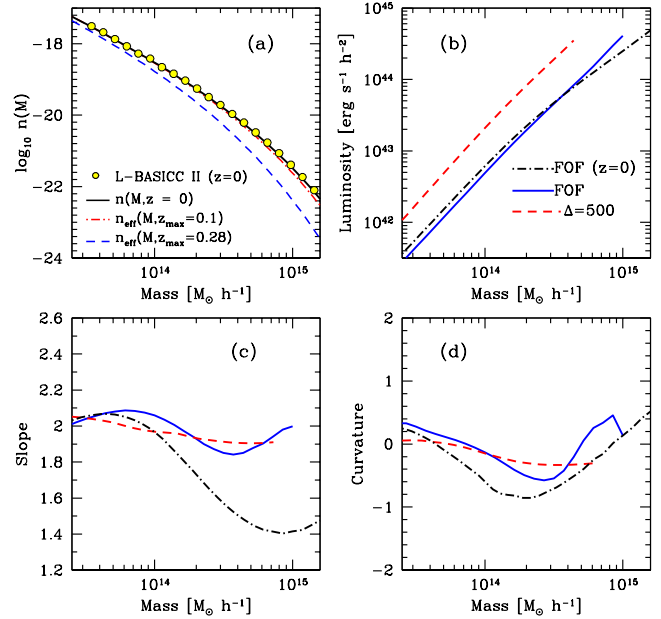}
\caption{(a) Mean halo mass function $n(M)$ (filled circles) of the L-BASICC II simulations at $z=0$. The solid line shows the prediction based on the fit of
\citet{jenkins}. The dotted and dashed lines show the effective mass function derived from a light-cone (following the REFLEX II sensitivity map (see \S~\ref{sec:phys})) for two different limiting X-ray luminosities. (b) $M$-$L_{X}$ relation obtained from the abundance matching technique using the mass function at $z=0$ (dot-dashed line), and light-cone estimates using the FOF (solid line) and SO haloes with $\Delta=500$ (see \S~\ref{sec:phys}). (c) Slope (first derivative) of the $M$-$L_{X}$ relation. (d) Curvature (second derivative) of the $M$-$L_{X}$ relation.}\label{abundance_2}
\end{figure}

Under the assumption of virial and hydrostatic equilibrium, constant gas mass to total mass ratio and thermal bremmstralhung being the dominant emission
mechanism in the intra-cluster medium \citep[see][for a review]{voit}, the $M$-$L_{X}$ relation should be well described by a power-law, at
least for the brightest galaxy clusters \citep[e.g][]{lm0,stanek,mau,rykoff,pratt,vik}. Furthermore, within the \emph{ROSAT} energy band, the X-ray emissivity
is weakly dependent on the temperature of the gas, specially for the hot clusters ($T\geq 3$ keV or $L_{X}\gtrsim L_{\star}$). These facts can be shown to
generate scaling relations of the form $L_{X}\propto M_{\rm vir}$ and $L^{\rm bol}_{X}\propto M_{\rm vir}^{4/3}$ for band and bolometric luminosities respectively.
Deviations from self-similar expectations are associated with non-thermal processes such as AGN feedback from the growth of super-massive black-holes \citep[e.g][]{borgani05,kay,puchwein,stanek_last,short}. These feedback processes lead to a more significant gas depletion for low mass systems, reducing the X-ray
luminosity and breaking the simple self-similarity as a function of mass. The evolution of the $M$-$L_{X}$ relation has been studied in observations
\citep[e.g][]{vik,reichert} and found to also show deviations from the simple self-similar model. 

A simple non-parametric estimate of the underlying $M$-$L_{X}$ relation can be obtained by implementing the so-called abundance matching technique \citep[e.g][]{behroozi},
in which we determine the X-ray luminosity $\bar{L}_{X}(M)$ as a function of the halo mass by matching the cumulative abundances $\int_{\bar{L}_{X}}^{\infty}\Phi(L')\dd L'=\int_{M}^{\infty}n(M')\dd M'$. This method is applicable under the assumption that X-ray luminosity is a monotonically increasing function of the halo mass with
no intrinsic scatter. The resulting $M$-$L_{X}$ relation is shown by the black dot-dashed line in panel (b) of Fig.~\ref{abundance_2}, displaying deviations
from a power-law. To obtain this result we used the recipe for the halo mass function of \citet{jenkins} computed for the cosmological model of the L-BASICC II
simulations at $z=0$.

In order to quantify the inferred departure from a power-law $M$-$L_{X}$ relation, the dot-dashed lines in panels (c) and (d) of Fig.~\ref{abundance_2} show the
behaviour of its first and second derivatives, respectively, as a function of the halo mass. Within the luminosity range explored by the REFLEX II X-ray luminosity
function, the slope of the $M$-$L_{X}$ relation displays a transition from a maximum value of approximately $\sim 2$ at $\sim 5\times 10^{13}M_{\odot} h^{-1}$ to
$\sim 1.4$ at a mass scale of $\sim 10^{15}M_{\odot}h^{-1}$. Accordingly, the curvature displays positive values in the low halo mass regime
$M\lesssim 10^{13.5} M_{\odot} h^{-1}$. In the intermediate halo mass range $5\times 10^{13} \lesssim M\lesssim 10^{15}M_{\odot}h^{-1}$, the curvature displays
negative values and resume positive values for the high-mass haloes. Although these results are based on a simple halo abundance matching technique which ignores
relevant physical effects, they suggest that a simple power-law relation 
$M$-$L_{X}$ will not be enough to describe the observed XLF through Eq.~(\ref{eq:flum_theo}).

We calibrate the $M$-$L_{X}$ relation that we use in the construction of our mock catalogues directly from the halo samples in the N-body simulations. 
We assume that X-ray luminosities are distributed around a mean scaling relation with a log-normal scatter in the luminosity (at a fixed mass)
$\sigma_{\ln L|M}$. In order to account for the deviation from a power-law as suggested by the abundance matching technique, we adopted a mass-dependent slope
(or running-index scaling-law) to parameterise the mean $M$-$L_{X}$ relation in the whole luminosity range covered by the REFLEX II sample. 
Defining $\bar{\ell} \equiv \log_{10} (\bar{L}_{X}/10^{44}{\rm erg}\,{\rm s}^{-1}\,h^{-2})$ and $m \equiv \log_{10} (M/10^{14}M_{\odot}h^{-1})$, our input mean
scaling relation $M$-$L_{X}$ has the form 
\be\label{m_lum_relation}
\bar{\ell}(m)=a+bm+cm^{2},
\ee
where $a$ determines the mean luminosity at $10^{14}M_{\odot}\,h^{-1}$, $b$ determines the slope at that mass-scale and the parameter $c$ characterises the curvature.
We assigned X-ray luminosities to the dark matter haloes using Eq.~(\ref{m_lum_relation}), applying a log-normal scatter around $\ln \bar{L}_{X}$
characterised by $\sigma_{\ln L|M}=0.26$ ($\sim 30$ per cent scatter). This value was derived by \citet{stanek} based on cluster
number counts and spatial clustering measured from a sub-sample of the REFLEX catalogue. We note however that such a value was determined under the assumption of
a power-law scaling relation. Furthermore, in their analysis, \citet{stanek} assumed a fiducial cosmological model with $\Omega_{\rm mat}=0.24$ and a power
spectrum normalisation of $\sigma_{8}=0.85$, which is higher than our fiducial value $\sigma_{8}=0.773$). \citet{tr06} verified that the high intrinsic
scatter measured by \citet{stanek} was related to the cosmological model they implemented, mainly, to the high value of $\sigma_{8}$. Using the cosmological
parameters derived from WMAP-3year data \citep{spergel07}, \citet{tr06} found that a small intrinsic scatter was still able to describe the observations (within
the mass uncertainties). \citet{mantzI} used a fiducial cosmological model similar to ours, and found a larger scatter ($\sim 40$ per cent), though it was 
substantially reduced ($\sim 6 $ per cent) when core-excised clusters were used in their analysis. We therefore emphasise that the parameters of the $M$-$L_{X}$
relation obtained with a fixed scatter and from a FOF halo mass function \emph{are only meant to reproduce the cluster abundance within the N-body simulations and $z=0$.}

On top of the intrinsic scatter, we introduced the errors in the X-ray luminosity through a second log-normal scatter with $\sigma=0.2$. These two scatters can be added in quadrature, such that the total  $M$-$L_{X}$ relation preserves a log-normal scatter given by
$\tilde{\sigma}=0.328$ ($39$ per cent scatter). After luminosities were assigned to each halo we determined the luminosity functions within the $50$ simulation boxes.
We implemented a MCMC algorithm to determine the set of parameters $(a,b,c)$ in Eq.~(\ref{m_lum_relation}) by comparing 
the obtained luminosity distributions with the measured REFLEX II XLF, assuming a Poisson-noise standard deviation, as pointed out in \S~\ref{subsec:xlf}.

As a consistency check, we have verified that replacing the measured XLF by the best fit extended-Schechter function of Eq.~(\ref{XLFe}) yields a 
set of parameters $(a,b,c)$ different by less than $5$ per cent.

\begin{figure}\includegraphics[width=8.5cm, angle=0]{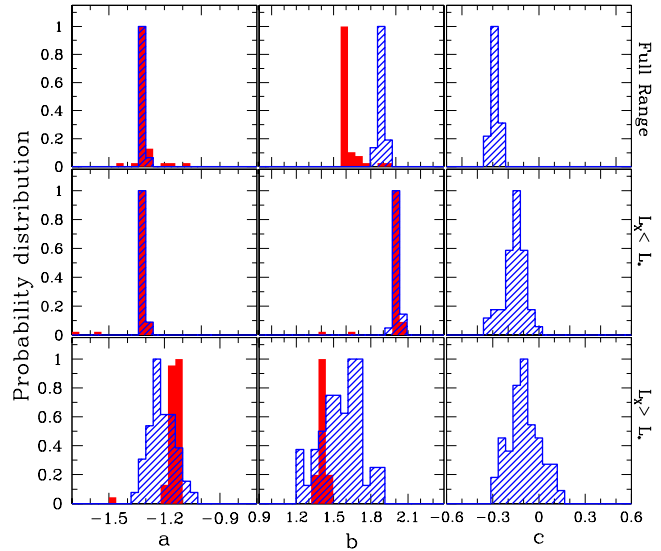}
\caption{Distribution of the best-fitting parameters $(a,b,c)$ determined from the $50$ realizations of the
L-BASICC II simulations within the full luminosity range as well as in the faint and bright tail of the XLF.
Solid line distributions show the results assuming a $M$-$L_{X}$ relation given by Eq.~(\ref{m_lum_relation}). Filled histograms show the distribution of parameters associated with a power-law ($c=0$).}\label{ml_distrib}
\end{figure}

\subsection{The inferred $M$-$L_{X}$ relation}\label{sec:results}

\begin{figure}
  \includegraphics[width=8cm, angle=0]{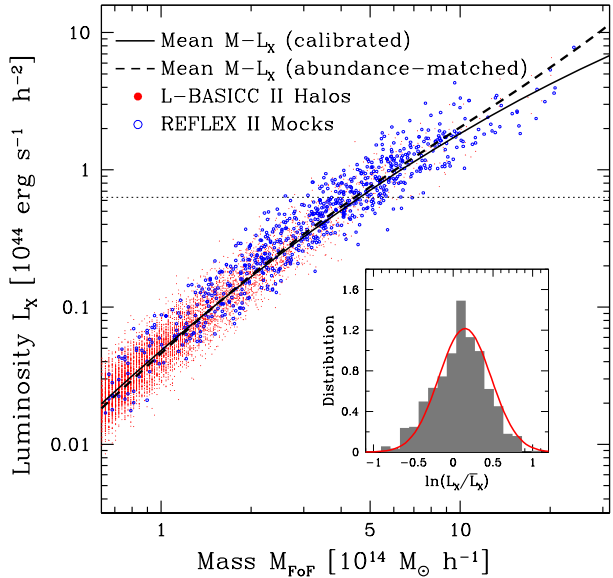}
  \caption{$M$-$L_{X}$ relation of the REFLEX II mock catalogues. The solid line represents the mean $M$-$L_{X}$
relation calibrated from $\Phi(L_X)$. Filled circles correspond to individual halos in one of the L-BASICC II 
simulations with X-ray luminosities assigned as described in \S~\ref{sec:ml}). Open circles correspond to mock clusters 
in one member of the ensemble of $100$ REFLEX II mock catalogue. The dashed line represents the
$M$-$L_{X}$ relation determined by the abundance matching technique (\S~\ref{sec:ml}). The horizontal line marks the value $L_{\star}$.
The inner plot shows the distribution of the luminosities with respect to their mean value. The solid line
represents a Gaussian distribution with dispersion $\tilde{\sigma}$ centred at $\ln(L/\bar{L})=(\dd \ln V/\dd \ln 
L)\tilde{\sigma}^{2}\sim 0.14$, where $V(L)$ is the volume as a function of the X-ray luminosity (see 
\S~\ref{sec:basic}).}\label{lm_relation}
\end{figure}

Following the procedure described in \S~\ref{sec:mla}, we obtained best-fitting values of the parameters 
$(a,b,c)$ from each of the 50 L-BASICC II realizations. The distributions of these values are shown in the upper
panels of Fig.~\ref{ml_distrib}, and present narrow peaks around the values $a=-1.36\pm0.03$, $b=1.88\pm0.05$  and $c=-0.29\pm0.04$ (where the errors represent
the variance from the ensemble of simulations). These values, together with the effective scatter $\tilde{\sigma}=0.328$, were used in the final construction
of the mock catalogues described in \S~\ref{sec:ml} and used in Paper I. The constraints on $c$ indicate a $\sim 7\sigma$ deviation from the power-law $M$-$L_{X}$ relation (i.e. $c=0$). The solid line in Fig.~\ref{lm_relation}
shows the mean $M$-$L_{X}$ relation obtained with these set of parameters.
The points correspond to individual halos in one of the L-BASICC II realizations with X-ray luminosities assigned following our $M$-$L_{X}$ relation, while
open circles represent clusters in one of our final REFLEX II mock catalogues (i.e, once the REFLEX II selection function is applied as described in \S~\ref{sec:ml}).
As shown by the solid histogram in Fig.~\ref{lcasos}, the luminosity distribution generated through this calibration provides an excellent match to the measured
X-ray luminosity function of the REFLEX II sample. 

The $M$-$L_{X}$ relation obtained through this calibration procedure is in agreement with that derived from the simple abundance matching technique described in
\S~\ref{subsec:xlf} (shown by the dashed line in Fig.~\ref{lm_relation}), especially for $M\lesssim 5\times 10^{14}M_{\odot}h^{-1}$ which corresponds to a luminosity $L_{X}\sim L_{\star}$. At higher masses, the difference between the two is clear. Due to the exponential fall-off of the halo mass function at such high mass scales, the obtained X-ray luminosity distribution of the halos is very sensitive to the assumed scatter in the $M$-$L_{X}$ relation.

As a consistency check, we have verified that the values $(a,b,c)$ obtained from the resulting distribution within the L-BASICC II simulations are in good agreement
with those obtained when the fitting formulae of \citet{jenkins} is used to compare the measured $\Phi(L_X)$ against the prediction from Eq.~(\ref{eq:flum_theo}). The
results obtained in this case are $a=-1.29\pm 0.04$, $b=1.95\pm 0.09$ and $c=-0.31\pm 0.06$, where the errors denote the $1\sigma$ confidence intervals. 

A power-law scaling relation could still be sufficient to reproduce the abundance of high luminosity ($L_{X}>L_{\star}$) galaxy clusters. In order to verify this, we repeated the procedure described in \S~\ref{sec:mla}
to calibrated the $M$-$L_{X}$ relation in the faint ($L_{X}\leq L_{\star}$) and bright ($L_{X}>L_{\star}$) tails of
the luminosity function. The results are shown in the second and third rows of Fig.\ref{ml_distrib}. The set of parameters calibrated in the first interval
($L_{X}<L_{\star}$) are marginally compatible with a power-law
(for instance $\sim 6$ per cent of the realizations displayed curvatures in the range $|c|\leq 0.05$), whereas the analysis derived from the bright tail is 
more consistent with such scaling relation ($\sim 25$ percent of the realizations had $|c|\leq 0.05$). We have checked that by pushing the limiting luminosity
up to higher values, the resulting distribution of the curvature is well peaked at the value $c\sim 0$, though the distribution becomes wider due to the low number
of objects with such luminosities. Fig.~\ref{ml_distrib} also shows the distribution of the parameters associated with a power-law $M$-$L_{X}$ relation, using the full
range as well as the faint and bright end tails of the luminosity distribution. The evident shifts in the peaking values of the slope $b$ in these two luminosity
ranges shows the need for a mass-dependent slope, or in general, for deviations from a power-law $M$-$L_{X}$ relation in case the full luminosity range is used in
the analysis. 

%\scr{As pointed out in \S~\ref{sec:sample}, we assumed that the flux errors (and therefore the errors in the X-ray luminosity) can be drawn from a normal distribution
%with a fixed width}. A calibration scheme in which the errors depend upon other quantities within the sample (as discussed in \S~\ref{sec:sample}) would be time demanding and would lead to a different set of parameters $(a,b,c)$, which, by construction, generate a luminosity function $\Phi(L_{X})$ that matches that of the REFLEX II sample. 

 \begin{table*}
\center
\begin{tabular}{l|l|c|c|c|} \hline \hline
& &$a$  &  $b$  & $c$ \\
&&\\
i  & FOF L-BASICC II  Simulation &$-1.36\pm0.03 $ &   $1.88\pm0.05$ & $-0.29\pm0.04$ \\
ii & FOF L-BASICC II  Simulation: masses corrected &$-1.10\pm 0.02$ &   $1.70\pm0.04$ & $-0.23\pm 0.04$ \\
iii& FOF \citep{jenkins} & $-1.29\pm 0.04$   & $1.95\pm 0.09$ & $-0.31\pm 0.06$\\

iv & SO $\Delta=500$ \citep{tinker} full $L_{X}$ range &$-0.68\pm 0.06$   & $1.65\pm 0.03$ & $-0.38\pm 0.04$\\ 
v  & SO $\Delta=500$ \citep{tinker} $L>1.6L_{\star}$ &$-0.69\pm 0.07$   & $1.45\pm 0.26$ & $-0.15\pm 0.22$\\ 
vi  & SO $\Delta=500$ \citep{tinker} $L>1.6L_{\star}$, power-law &$-0.64\pm 0.03$   & $1.27\pm 0.05$ & $0$\\  \hline \hline
\end{tabular}\caption{Parameters of the $M$-$L_{X}$ relation in Eq.~(\ref{m_lum_relation}) obtained by calibration against the observed REFLEX II XLF: i) using the L-BASICC II simulations, ii) using the L-BASICC II simulations with FOF masses corrected as \citet{more}, iii) using the \citet{jenkins} fitting formula for the halo mass function, and iv) \& v) using the \citet{tinker} mass function for SO masses with $\Delta=500$ in two luminosity intervals. In order to compare with the analysis of \citet{mantzI}, the last row vi) shows the results obtained by fitting a power-law in the range $L>1.6L_{\star}$.}
\label{table}
\end{table*}

\subsection{On the shape of the  $M$-$L_{X}$ scaling relation}\label{sec:phys}
As discussed in \S~\ref{sec:data}, the determination of the X-ray luminosities of the REFLEX II sample assumed, 
among other things, a fiducial set of cosmological parameters and a definition of halo masses, namely, SO masses
with  $\Delta=500$. In this subsection we briefly explore the possible impact of the differences between these
fiducial values and the corresponding quantities characterising the L-BASICC II N-body simulations, on the
recovered values of the parameters of the $M$-$L_{X}$ relation.

\begin{itemize}
\item \emph{Resolution effects}. It has been shown \citep[e.g][]{warren,lukic,more} that the finite number of particles defining a dark matter halo is overestimated by the FOF finder algorithm, when compared to the same masses obtained with simulations of higher resolution. In particular, the empirical correction of \citet{warren}, based on the number of particles defining a dark matter halo, was designed to match the abundance of low resolution to that of higher resolution simulations. This correction leads to a $40$ per cent mass overestimation at the resolution limit of the L-BASICC II simulations, reducing to a few percent at $10^{15}M_{\odot}\,h^{-1}$. Based on percolation theory, this systematic effect has been modeled by \citet{more}, who showed that the FOF masses are overestimated by $\sim 30$ per cent at the resolution of the L-BASICC II simulations, and by $15$ per cent at a mass scale of $10^{15}M_{\odot} h^{-1}$ (S. More, private communication), representing a higher correction than that of \citet{warren}. In order to check the impact of this systematic effect on our results, we corrected the halo masses within the L-BASICC II simulations and re-calibrated the scaling relation given by Eq.~(\ref{m_lum_relation}) on a number of realizations. The results show a $15$ per cent increment in the normalization $a$, with its corresponding decrease the slope $b$ (around $10$ per cent) and a shift of the curvature $c$ towards positive values (around $15$ per cent), with respect to our original estimations. In particular, the shift in the curvature corresponds to approximately $1\sigma$, excluding the value $c=0$ with a statistical significance of $\sim 4\sigma$. We can conclude that the correction for the finite number of particles in some extent diminishes the departure from a power-law scaling relation, although not completely.  
%In this concern, we note that the amplitude of the halo mass function after correcting the halo masses is $\sim 40$ per cent smaller than that of the uncorrected mass function. The difference between these two measurements is, to a good approximation, constant within the mass range of the L-BASICC II. Therefore, had we applied the abundance-matching technique with a fitting formula based on \emph{corrected halo mass function}, the magnitude of the differences between the resulting parameters of the scaling relation with respect to those obtained from the \emph{uncorrected halo mas function} would have been higher for the normalization $a$, followed by those in the slope $b$ and the curvature $c$.

\item \emph{Cosmological parameters.} We have tested the impact of the cosmological parameters in the calibration of
the scaling relation, by constraining the set $(a,b,c)$ using the \citet{jenkins} mass function at $z=0$ computed
for different values of $\Omega_{\rm mat}$ in the range $0.15-0.35$, keeping both $\sigma_{8}$ fixed to our fiducial value or letting it vary as $\sigma_{8}=0.45\Omega_{\rm mat}^{-0.3}$ \citep[e.g][]{evrard}. In both cases, while the amplitude $a$ (the slope $b$) decreases (increases) with respect to $\Omega_{\rm mat}$, the curvature $c$
remains approximately constant, $c\neq0$ with a $2\sigma$ significance.

\item \emph{Redshift evolution}. The redshift evolution of the mean matter density of the Universe can be imprinted in the amplitude of the $M$-$L_{X}$  relation \citep[e.g][]{borgani1,voit}. The parameterisation given by Eq.~(\ref{m_lum_relation}) can then be thought of as accounting for such evolution through the conversion $A(z)\to 10^{a}M^{cm}$, in case the true underlying $M$-$L_{X}$ relation is given by a power-law $L_{X}=A(z)M^{b}$. On the other hand, even without a redshift dependent amplitude
(i.e, $A(z)$ constant), the function $p(L_{X}|M,z)$ could still evolve with redshift through its intrinsic scatter \citep[e.g][]{fabjian}. For the purposes of building mock catalogues
from a snapshot of the simulations, such dependency could be translated to a mass-dependent scatter. We explored this possibility by calibrating a power-law $M$-$L_{X}$
relation with a scatter given by $\sigma(M)=\sigma_{\rm ln L|M}(M/10^{14}M_{\odot}h^{-1})^{\gamma}$. The resulting X-ray luminosity distribution generated by this
parameterisation (with best-fitting values $a\sim -1.2$, $b\sim 1.8$ and $\gamma\sim -2.3$ with $\sim 50$ per cent error in $\gamma$) is shown by the dot-dashed histogram in Fig.~\ref{lcasos}. Such parameterisation
represents a mild improvement with respect to the power-law or running index $M$-$L_{X}$ relation in the bright end ($L>L_{\star}$) of the XLF, while working as good as the power-law for $L\lesssim L_{\star}$ clusters.

\item \emph{Light-cone effects}. In order to take into account the redshift evolution of the halo abundance, the XLF measured within a past light-cone must be compared with the volume average of Eq.~(\ref{eq:flum_theo}). Under the assumption of a redshift-independent $M$-$L_{X}$ relation, we can write the prediction for the luminosity functions
in the light-cone as
\be
\Phi(L_{X})=\int_{0}^{\infty}  n_{\rm eff}(M,z_{\rm max})p(L_{X}|M)\dd M,
\ee
where $n_{\rm eff}(M,z_{\rm max})$ is the halo mass function averaged over the volume of the light-cone defined by the sample:
\be
\label{eq:neff}
n_{\rm eff}(M,z_{\rm max})=\frac{1}{V(z_{\rm max})}\int_{V(z_{\rm max})} n(M,z)\frac{\dd V}{\dd z} \dd z,
\ee
with $z_{\rm max}=z_{\rm max}(L_{X})$ as the maximum redshift allowed for a cluster with luminosity $L_{X}$ in order to be detected, and
$V(z_{\rm max})$ is the comoving volume at that redshift. This information is encoded in the REFLEX II sensitivity map described in \S~\ref{sec:sample}. As an example,
we show the behaviour of $n_{\rm eff}(M,z_{\rm max})$ in panel (a) of Fig.~\ref{abundance_2}, based on the halo mass function for FOF and SO masses. In order to apply the
match abundance technique to derive an estimate of the shape of the  $M$-$L_{X}$ relation using Eq.~(\ref{eq:neff}), we carried out an iterative process in
which an initial power-law scaling relation was assumed, $\ell=-1.0+1.5\,m$, from which $n_{\rm eff}(M,z_{\rm max}(M))$ could be determined in order to match the abundances, leading
to a new estimation of the $M$-$L_{X}$ relation, which is used as an input for the next iteration. The resulting $M$-$L_{X}$ relations (convergence is achieved
after few iterations) derived from the FOF and SO mass functions are shown in panel (b) of Fig.~\ref{abundance_2}, with its first and second derivatives in panels
(c) and (d) respectively. The recovered scaling relation still displays departures from a power-law,  although the curvature is slightly smaller than that
derived from the simulation snapshots (dot-dashed line in Fig.~\ref{abundance_2}). The deviation from a power-law is more evident for FOF masses than for SO masses. This might indicate that the deviations from a power-law scaling relation
observed by matching the abundances at redshift $z=0$ are not significantly accounted for by the evolution of the halo-abundance within the light cone defined by the
REFLEX II sample, at least in the case of vanishing intrinsic scatter. 

\item \emph{Definition of halo mass}. We calibrated the $M$-$L_{X}$ relation given by Eq.~(\ref{m_lum_relation}) using the fitting formulae of \citet{tinker} for SO haloes with $\Delta=500$. The mass function was evaluated at the median redshift of the sample $\bar{z}\approx 0.09$. Fixing the intrinsic scatter to our fiducial value we obtained $a=-0.68\pm 0.06$,  $b=1.65\pm 0.03$ and $c=-0.38\pm 0.04$. The value for the curvature excludes the power-law scaling relation with $\sim 7\sigma$ confidence. Similarly, for clusters with $L_{X}>L_{\star}$ (with a median redshift of $\bar{z}\approx 0.14$) we found $a=-0.73\pm 0.02$, $b=1.68\pm 0.16$, and $c=-0.34\pm 0.14$, excluding the power-law behavior to $\sim 2\sigma$ significance. We calibrated a power-law scaling relation in the range $L_{X}>1.6L_{\star}\approx 10^{44}{\rm erg}\,s^{-1}\,h^{-2}$, similar to that used by \citet{mantzI}, obtaining $a=-0.64\pm 0.05$ and $b=1.27\pm 0.05$, in good agreement with that analysis. In the same range, the fit to Eq.~(\ref{m_lum_relation}) yields a curvature of $c=-0.15\pm 0.23$, compatible with a power-law.

\end{itemize}
The tests presented in this section show that the deviation from a power-law scaling relation in the full X-ray luminosity range probed by the REFLEX II XLF is likely not due to
differences in the mass definition, the cosmological parameters or evolution along the light cone. This is nevertheless not conclusive, in the sense that a full
analysis including the evolution of the scaling relation, its intrinsic scatter and the uncertainties in the halo abundance is still required in order to draw
stronger conclusions about the cluster scaling relations.

Table~\ref{table} summarizes the values of the parameters $(a,b,c)$ obtained from the various analysis presented in this section.

\begin{figure}
\includegraphics[width=8cm,angle=0]{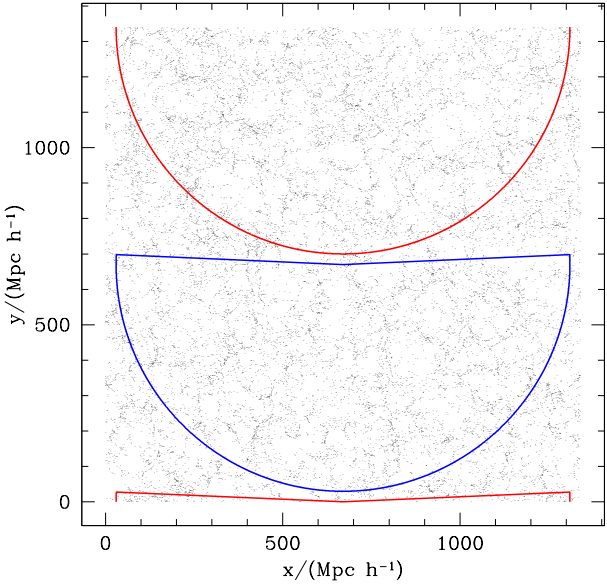}
\caption{Slice of $100\,{\rm Mpc}\,h^{-1}$ width through the centre of the halo distribution in one of the
L-BASICC II simulations. The solid lines indicate the geometric configuration used to extract two independent
mock catalogues with $z_{\rm max}=0.22$ from each realization.}
\label{slice}
\end{figure}

\subsection{The construction of the mock catalogues}
\label{sec:ml}

We now describe the procedure followed to construct the REFLEX II mock catalogues. 
First, we note that in Paper I it was shown that the effective
volume for clustering analyses \citep[e.g][]{FKP} probed by the REFLEX II sample does not increase
substantially beyond redshift $z\approx 0.22$. As the primary goal of the set of mock catalogues is to determine
covariance matrices for the statistical analysis of the large-scale structure probed by the REFLEX II sample,
we set this value as the maximum redshift of the mock catalogues. This in turn allowed us to construct $100$
independent mocks catalogues out of the $50$ L-BASICC II realizations. Mock catalogues with higher 
maximum redshifts can in principle be constructed, at the cost of having less independent realizations.

The geometric arrangement of the mock catalogues is depicted in Fig.~\ref{slice}, where a slice cut of 
$100\,h^{-1}{\rm Mpc}$ through the center of one of the L-BASICC II realizations is shown. Given such geometric arrangement, we assigned a cosmological redshift $z_{\rm cos}$ to each halo by inverting the redshift-comoving
distance relation $r(z_{\rm cos})=\int_{0}^{z_{\rm cos}} \dd z' H^{-1}(z')$ for the cosmology of the simulations. 
%Next we transformed
%the coordinates of the haloes to redshift space via $r\to r+\textbf{v}\cdot\hat{\textbf{r}}/H_{0}$,
%where $\textbf{v}$ is the peculiar velocity of the centre of mass of the haloes.
The final redshifts of the halos were obtained taking into account the distortions 
introduced by their peculiar velocities $v$ as $z=(1+z_{\rm cos})(1+z_{\rm pec})-1$, where $z_{\rm pec}=v/H_{0}$.
For the $z=0$ output of the L-BASICC II simulations, the distribution of peculiar velocities displays a
mean value of $\bar{v}\approx 350\,{\rm km}\,{\rm s}^{-1}$ with a velocity dispersion of 
$\sigma\approx 180\,{\rm km}\,{\rm s}^{-1}$, which in terms of redshift corresponds to
$\Delta z \sim6\times 10^{-4}$. 
As can be seen in table 2 of \citet{guzzo_optical}, the redshift distortions induced by spectroscopic redshift uncertainties in the REFLEX II sample are significantly smaller than those induced by peculiar velocities. Therefore we do not introduce such effect in the estimate of the redshift of our mock clusters.

We assigned X-ray luminosities to the dark matter haloes according to the  $M$-$L_{X}$ relation described
in \S~\ref{sec:results}. These halos are then \emph{observed} through the REFLEX II sensitivity map, rejecting
those haloes with luminosities below the limiting luminosity $L^{\rm lim}_{X}$ at a given location.
The open circles in Fig.~\ref{lm_relation} correspond to individual objects in one of our final REFLEX II mock catalogues which,
with a minimum mass $M\sim 3.5\times 10^{13} h^{-1}M_{\odot}$, corresponding to a 20 particle dark matter halo,
covers a mass range well within the resolution limit of the simulations.

\begin{figure}
\includegraphics[width=8cm,angle=0]{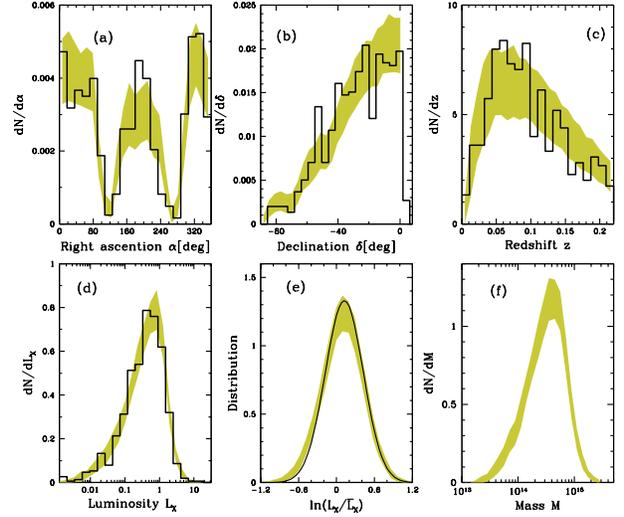}
\caption{Properties of the mock catalogues: (a) right-ascension, (b) declination, (c) redshift (d) luminosity (e) luminosity relative to the true underlying $M$-$L_{X}$ relation and (f) mass distribution. The shaded areas show the standard deviation in the distribution of each property derived from the set of mock catalogues and compared to the corresponding property in the REFLEX II sample (solid line). In panel (f) the solid line shows a Gaussian distribution centred at $\delta \tilde{\sigma}^{2}$ (see \S~\ref{sec:basic})}\label{fig_prop3}
\end{figure}

\section{Properties of the REFLEX II mock catalogues}
\label{sec_properties}

In this section we discuss some of the most relevant properties of the REFLEX II mock catalogues.

\subsection{Basic properties}
\label{sec:basic}

The underlying $M$-$L_{X}$ relation applied to construct the REFLEX II mock catalogues (and used in the clustering analysis of Paper I) is characterised by Eq.~(\ref{m_lum_relation}), with the set of parameters shown in \S~\ref{sec:results}. The performance of the REFLEX II sky-mask is shown in panels (a) and (b) of Fig.~\ref{fig_prop3} by means of the distribution in equatorial coordinates of the mock catalogues, compared to that of the REFLEX II sample, shown by the solid line. The Milky-way band is well described in panel (a) by the two minimum distribution values at $\alpha\sim 100$ and $\alpha\sim 250$ deg (see also fig. 1 of Paper I). Panels (c) and (d) show respectively the
redshift and X-ray luminosity distributions, displaying good agreement with that of the real REFLEX II sample. 
The mass distribution within the mock catalogues is shown in panel (f) of Fig.~\ref{fig_prop3}. 
This implies an average halo mass of $M\simeq4\times 10^{14} h^{-1}M_{\odot}$ for the REFLEX II clusters.

%$\sim 3.5\times 10^{13} h^{-1}M_{\odot}$, corresponding to dark matter haloes with $\gtrsim 20$ particles. The most massive cluster has a mass of $\sim 5 \times 10^{15}M_{\odot} h^{-1}$. 

\begin{figure}
\includegraphics[width=8cm]{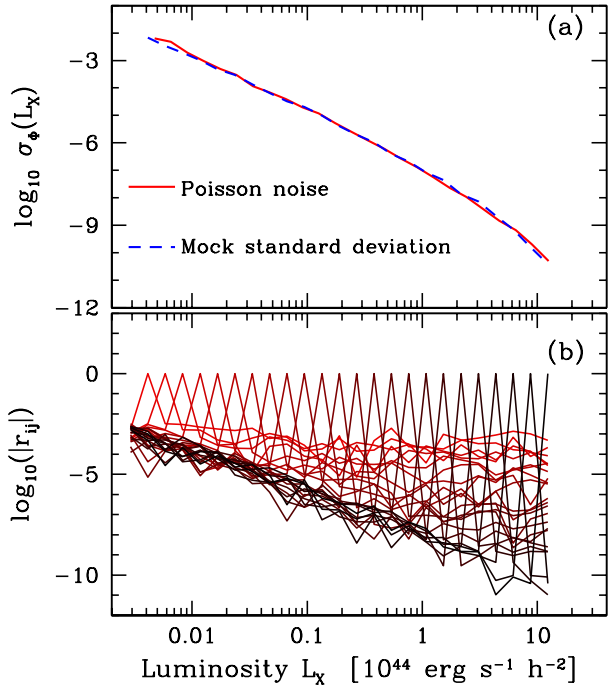}
\caption{
Covariance matrix of the XLF. Panel (a) shows the comparison between the standard deviation of the XLF derived from the set of mock catalogues, and the Poisson-noise prediction determined from the number of clusters in each bin of X-ray luminosity. Panel (b) shows the correlation coefficients $r_{ij}$ of the XLF.}\label{cova} 
\end{figure}

The application of the REFLEX II selection function to the N-body simulations, introduces in our set of mock
catalogues a selection effect inherent to all flux-limited samples, namely, the fact that objects with 
increasing luminosities probe larger cosmological volumes. In combination with the intrinsic scatter in the
$M$-$L_{X}$ relation this introduces the so-called Malmquist bias which, if unaccounted for, can lead to
biased estimates of the underlying scaling relation, especially when a threshold in luminosity is imposed to
the sample. Our set of mock catalogues can then be used as a control sample to test different methods
to obtain scaling relations taking into account the selection effects mentioned above. At a given mass $M$ the Malmquist bias causes the
mean observed $M$-$L_{X}$ relation $\langle L(M) \rangle $ to deviate from the underlying $M$-$L_{X}$ relation
$\bar{L}(M)$ according to \citep[e.g][]{stanek,vik,pratt}
\begin{equation}
\langle \ln L_{X}\rangle=\ln \bar{L}_{X}(M) + \delta\tilde{\sigma}^{2},
\end{equation}
where it is assumed that the volume probed by a cluster of luminosity $L_{X}$ scales as
$V(L_{X})\propto L_{X}^{\delta}$ ($\delta=0$ for a volume limited sample) together with a
log-normal distribution function $p(L_{X}|M,z)$ with scatter $\tilde{\sigma}$. 
Under the assumption of a non-evolving $M$-$L_{X}$ relation, the Malmquist bias can be
translated into an overestimation of the slope and an underestimation of the amplitude in the
$M$-$L_{X}$ relation \citep[e.g][]{lm0}.
In panel (e) of Fig.~\ref{fig_prop3} we show the distribution of
luminosities around the mean $M$-$L_{X}$ relation given by
Eq.~(\ref{m_lum_relation}), for the $100$ mock catalogues. The dashed
line represents a log-normal distribution centred at $\ln (L/\bar{L})=\delta
\tilde{\sigma}^{2}\approx 0.14$, where we have used $\delta=1.38$, which gives
a good description of the maximum volume as a function of the X-ray
luminosity. In terms of luminosities, this bias is translated into a factor
$\sim 1.15$.

We finally study the behavior of the covariance matrix of the REFLEX II XLF. In panel (a) of Fig.~\ref{cova} we compare the standard deviation obtained from the final set of mock catalogues with that derived from a Poisson noise. The good agreement between these two estimates implies that the inhomogeneous spatial distribution of X-ray galaxy clusters (i.e, the sampling variance) has a minor impact on the determination of the variance of the XLF \citep[e.g][]{smith_xlf} given the comoving volume and luminosity range probed by the REFLEX II sample. Furthermore, we verified that the off-diagonal elements of the covariance matrix of the XLF are negligible compared to the variance. In this regard, panel (b) of Fig.~\ref{cova} shows the correlation coefficients defined by $|r_{ij}|\equiv |C^{\Phi}_{ij}|/\sqrt{C^{\Phi}_{ii}C^{\Phi}_{jj}}$ determined from the suite of mock catalogues, showing in general weak correlations between different luminosity bins, ranging from  $r_{ij}\sim 10^{-3}$ at low luminosities to $r_{ij}\sim 10^{-10}$ at high luminosities A more detailed analysis of the REFLEX II XLF will be presented in a forthcoming paper (B\"ohringer et al., in preparation).

\subsection{Clustering properties}

The power spectra of the REFLEX II mock catalogues were analysed in Paper I.
These measurements were obtained following the method described in Paper I, based on
the minimal-variance estimator of \citet{FKP}. This study showed that the mean power
spectra from our ensemble of mocks catalogues for different limiting luminosities is in excellent
agreement with those of the REFLEX II sample on intermediate and large scales
($0.02<k / (h\, {\rm Mpc}^{-1})<0.3$). 
While a detailed analysis of the clustering in configuration space will be presented in a forthcoming paper
(S\'anchez et al., in preparation), also in this case we find good agreement between the clustering signal 
of the REFLEX II sample and that of the mock catalogues.
This allows us to use our mock catalogues for its main application in the context of large scale structure,
that is, to obtain estimates of the covariance matrices of these measurements.

\begin{figure}
  \includegraphics[width=8.5cm, angle=0]{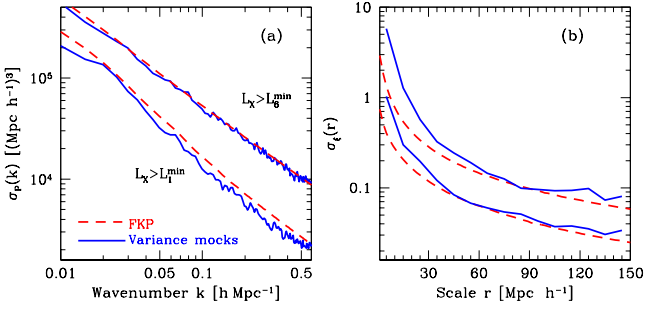}
  \caption{Variance of the power spectrum (panel a) and the correlation function (panel b) measured from the mock catalogues (solid lines) compared with the prediction from the estimator of \citet{FKP} (dashed lines), for two
  values of minimum X-ray luminosity.}\label{power_prop}
\end{figure}

Paper I presented an analysis of the bin-averaged covariance matrix $\hat{C}(k_{i},k_{j})$ of the cluster 
power spectrum. This covariance matrix showed highly correlated modes (with correlation coefficients $r_{ij}>0.9$) on intermediate ($k\sim 0.08 h\,{\rm Mpc}^{-1}$) and small scales ($k\sim 0.3 h\,{\rm Mpc}^{-1}$), mostly due to mode coupling induced by the survey window function. 
\citet{FKP} derived an approximated expression for the variance of the spherically averaged power spectrum under the assumption of Gaussian-distributed power around the true underlying value, given by: 
\begin{equation}
\frac{\sigma_{P}^{2}(k)}{P(k)^{2}}=\frac{2}{V_{k}V_{\rm eff}(k)},
\label{eq:sigmap}
\end{equation}
where $V_{k}\approx 4\pi k^{2}\delta k/(2\pi)^{3}$ is the volume of a spherical shell of width $\delta k$ and
$V_{\rm eff}(k)$ is the effective (coherence) volume probed by the survey at a scale $k$ \citep[e.g][]{FKP}. In its 
simpler form (for a volume limited sample), this contains the contribution from cosmic variance and shot-noise. 
The prediction of Eq.~(\ref{eq:sigmap}) can easily be translated into the variance of the two-point correlation function as \citep{cohn06,smith08,ariel_bao}:
\begin{equation}
 \sigma_{\xi}^2(r)=\frac{1}{2\pi^2}\int_0^\infty {\rm d}k\,k^2j_0(kr)^2\sigma_{P}^{2}(k),
\label{eq:sigmaxi}
\end{equation}
where $j_0(y)$ is the zero-th order spherical Bessel function.

Panel (a) of Fig.~\ref{power_prop} compares the variance of the cluster power spectrum derived from the set
of mock catalogues using sub-samples with minimum luminosities $L_1^{\rm min}=0.049$ and $L_6^{\rm min}=0.588\times 10^{44}
{\rm erg}\,{\rm s}^{-1}\,h^{-2}$ against the prediction of Eq.~(\ref{eq:sigmap}).
Similarly, panel (b) of the same figure shows the behaviour of the variance of the cluster correlation function
for the same values of minimum X-ray luminosity. Despite its simplicity, Eq.~(\ref{eq:sigmap}) gives a good 
account of the variances inferred from our mock catalogues. However, significant differences exist, especially for
the small-scale $\sigma_{\xi}$, which highlight the importance of the ensemble of mock catalogues as a
tool to obtain accurate estimations of covariance matrices which take into account the full effect of
non-linearities.

\begin{figure}
  \includegraphics[width=8cm, angle=0]{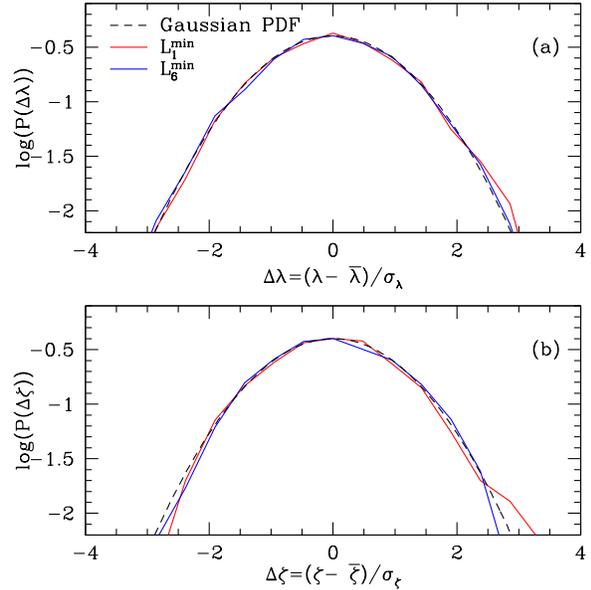}
  \caption{Probability distribution function (PDF) of the values of the power spectra (panel a) and correlation functions (panel b) in our ensemble of mock catalogues for two values of minimum X-ray luminosity. The dashed lines represent a Gaussian PDF.}
  \label{gaussian_pdf}
\end{figure}

Our ensemble of mock catalogues allows us to test another common assumption when comparing clustering measurements
against the predictions of cosmological models, that of a Gaussian likelihood function.
By studying the distribution function of the values of $P(k)$ and $\xi(s)$ obtained from the mocks
it is possible to reconstruct the shape of the likelihood function and find deviations from the simple Gaussian
assumption. As the covariance matrices for these statistics are not diagonal, this analysis is simplified by 
working on their ``de-correlated'' versions $\lambda_i$ and $\zeta_i$ defined as:
\begin{eqnarray}
 \lambda_i = (M_{P}^{-1})_{ij}P_j,\\
 \zeta_i = (M_{\xi}^{-1})_{ij}\xi_j,
\end{eqnarray}
where $M_{P}$ and $M_{\xi}$ are the transformation matrices of the basis where the covariance matrices $C_P$ and
$C_{\xi}$ are diagonal (given by their eigenvectors).  
Fig.~\ref{gaussian_pdf} shows the probability distribution function of 
$\Delta\lambda= (\lambda_i - \bar{\lambda}_i)/\sigma_i$ and 
$\Delta\zeta= (\zeta_i - \bar{\zeta}_i)/\sigma_i$ for two limiting luminosities for the power spectrum (panel a) 
and correlation function (panel b). These are well described by the prediction for a Gaussian distribution,
shown by the dashed line, indicating that deviations from the Gaussian case are very small.

\begin{figure}
  \includegraphics[width=8.5cm, angle=0]{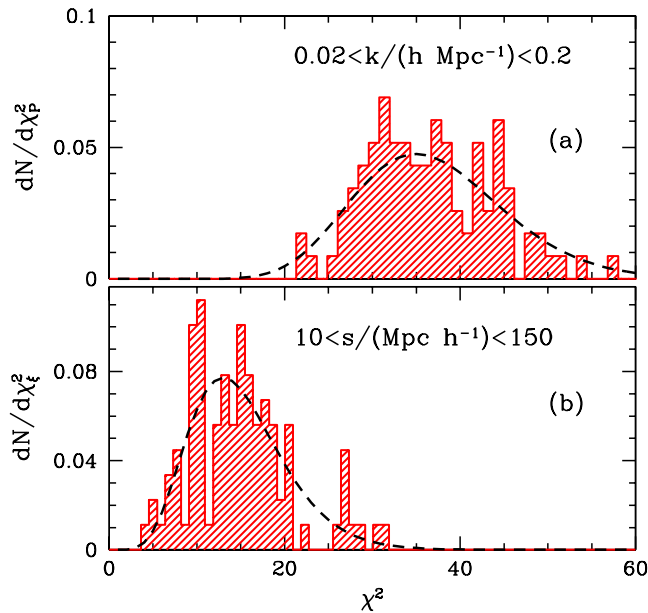}
   \caption{Distribution of the $\chi^{2}$ values associated to the power spectrum (panel (a)) and the correlation function (panel (b)) of the $100$ mock catalogues computed with respect to the mean of the ensemble. The dashed line represents the $\chi^{2}$ distribution for the corresponding number of modes in each interval of scales or wavenumbers.}\label{power_chis}
\end{figure}

Fig.~\ref{power_chis} shows the distribution of the chi-squared values $\chi^{2}$ associated with the cluster
power spectrum and the correlation function within the $100$ mock catalogues, computed with respect to their mean values within the ensemble.
The observed distributions are well described by a $\chi^{2}$ distribution \citep{abram} shown by the dashed lines, computed using the number of modes in each scale interval shown in the figure. 
This implies that the distribution of these quantities at a fixed scale is likely to follow a Gaussian distribution. We have tested this by performing a Kolmogorov-Smirnov test \citep[e.g][]{numerical_rec}, assuming as a null hypothesis a Gaussian distribution for the quantity 
$(\hat{P}(k_{i})-\bar{P}(k_{i}))/\bar{P}(k_{i})$, with zero mean and dispersion $\sigma(k_{i})/\bar{P}(k_{i})$. The test shows that on scales $0.02<k_{i}/(h\,{\rm Mpc}^{-1})\lesssim 0.2$ the distribution is compatible with the null-hypothesis to a $\gtrsim 88$ per cent level. We can therefore conclude that the resulting covariance matrix of the REFLEX II power spectrum can be used in an error analysis based on the assumption of Gaussian likelihoods. Similar conclusions can be obtained with respect to the clustering in configuration space.

\section{Conclusions}
\label{sec_conclusions}

In this paper we describe the procedure to construct a suite of galaxy cluster mock catalogues based on the X-ray luminosity function $\Phi(L_{X})$ of the REFLEX II sample and dark matter halo samples identified in the
L-BASICC II N-body simulations. 

The construction of our mock catalogues relies on the scaling relation between halo masses and 
X-ray luminosity in the \emph{ROSAT} energy band. Given the fact that the abundance of galaxy clusters is highly sensitive to the underlying scaling relation, we calibrate our mass proxy based on the X-ray luminosity
function of the REFLEX II sample and the halo-abundance of the N-body simulations. We find that in the
luminosity range probed by the REFLEX II sample, the observed $\Phi(L_X)$ demands a $M$-$L_{X}$ relation that
deviates from a power-law.

We qualitatively discarded such behaviour as a by-product of the differences between the features of the N-body simulations and the assumptions under which the REFLEX II catalogue was built, namely, the definition of halo
masses and the cosmological parameters. Similarly, the deviations from a power-law are likely not associated with possible light-cone effects (at least under the assumption of negligible intrinsic scatter). However, we find that a power-law is still sufficient to generate the observed abundance for bright clusters
($L_{X}\geq 10^{44}{\rm erg}\, s\,h^{-2}$), in agreement with recent results \citep[e.g][]{pratt,mantzI}. This might suggest that the observed break in the slope of the $M$-$L_{X}$ relation is a direct consequence of the physical properties of the galaxy clusters within the REFLEX II sample. A thorough analysis taking into
account both the evolution of the halo mass function and the evolution of the full $M$-$L_{X}$ relation is
required in order to draw more solid conclusions about the shape of the scaling relation as demanded from the
observed abundance.

%We qualitatively discarded such behaviour as a by-product of the differences between the features of the N-body simulations and the assumptions under which the REFLEX II catalogue was built, namely, the definition of halo
%masses and the cosmological parameters. Similarly, the deviations from a power-law are likely not associated with possible light-cone effects (at least under the assumption of negligible %intrinsic scatter). However, we find that a power-law is still sufficient to generate the observed abundance for bright clusters
%($L_{X}\geq L_{\star}$), in agreement with recent results \citep[e.g][]{pratt,mantzI}, while clusters populating the faint end of the X-ray luminosity function still demand deviations fro%m a power-law in the $M$-$L_{X}$ relation.
%This might suggest that the observed break in the slope of the $M$-$L_{X}$ relation is a direct consequence of
%the physical properties of the galaxy clusters within the REFLEX II sample. A thorough analysis taking into
%account both the evolution of the halo mass function and the evolution of the full $M$-$L_{X}$ relation is
%required in order to draw more solid conclusions about the shape of the scaling relation as demanded from the
%observed abundance.

Our set of mock catalogues reproduce the observed abundance of the REFLEX II sample as well as the observed
two-point clustering statistics of this data-set. Therefore, they provide an excellent tool 
to the test the statistical methods applied to the real data and to perform error analysis,
especially through the covariance matrix of the cluster power spectra or the correlation function.
The covariance matrix derived from this suite of catalogues properly accounts for physical effects such as
the non-linear evolution of clustering, its connection to the underlying scaling relations, as well as the
effect of the survey window function. This represents a major advantage of our set of mock catalogues against theoretical estimations of covariance matrices or log-normal catalogues. Our results also show that the distribution function of the values of $P(k)$ and $\xi(s)$ obtained
from the individual mocks are well described by the prediction for a Gaussian distribution.
This indicates that it is possible to assume a Gaussian likelihood function when comparing measurements of these statistics with the predictions
from cosmological models.

In the coming years, the next generation of galaxy cluster surveys
will probe much larger volumes than any present-day sample. 
For example, the \emph{eROSITA} cluster survey is expected to contain approximately $10^{5}$ objects (with $\sim 10^{3}$ with measured redshifts) \citep[e.g][]{pillepich} spanning a wide range of 
masses over a volume of $\sim 10^{2}\,{\rm Gpc}^3$. The construction of the mock catalogues necessary for the analysis of the large-scale clustering of this sample
will demand a great effort. This task will require a suite of extremely large N-body
simulations with a unique combination of volume and resolution to construct mock catalogues in the form
of light-cones up to a redshift $\sim1.5$.
The implementation of baryonic physics in such simulations would be even more ambitious and these mock catalogues
will need to rely on more indirect methods like the one discussed here.
Although the volume of the REFLEX II sample allowed us to implement a few simplifications such as the use of a
single snapshot of the simulations, the scheme we have implemented to calibrate the
$M$-$L_{X}$ relation can be extended to light-cones spanning wider redshift ranges, providing a simple technique 
to construct mock catalogues out of pure dark matter simulations even for these large samples. 

\section*{Acknowledgments}
We thank our anonymous referee for helping suggestions and comments that greatly improved the quality of the paper. We thank Ra\'ul Angulo and Carlton Baugh for providing us with the L-BASICC II simulations. We also thank Sarhud More for providing us with the correction for resolution effects in the FOF masses adapted to the L-BASICC II simulations. This research was in part supported by the DFG cluster of excellence ``Origin and Structure of the Universe''. CAC acknowledges support from STFC. This publication contains observational data obtained at the ESO La Silla observatory. 

\bibliographystyle{mn2e}

\end{document}